\documentclass[a4paper]{article}

\usepackage[english]{babel}
\usepackage[utf8x]{inputenc}
\usepackage[T1]{fontenc}

\usepackage[a4paper,top=3cm,bottom=2cm,left=3cm,right=3cm,marginparwidth=1.75cm]{geometry}

\usepackage{amsmath,amssymb,tikz-cd}
\usepackage{graphicx}
\usepackage{url}
\usepackage[colorinlistoftodos]{todonotes}
\usepackage[colorlinks=true, allcolors=blue]{hyperref}
\usepackage{authblk}
\usepackage{xcolor}
\usepackage{mathrsfs}
\usepackage[english]{babel}

\numberwithin{equation}{subsection}

\title{\textbf{Test gravitational waves in sandwich wave background}}
\author[1]{Ziqian Tang \thanks{tangziqian@pku.edu.cn}}
\affil[1]{School of Physics, Peking University, Beijing 100871, P. R. China}

\begin{document}
\maketitle
\begin{abstract}
The scattering of test fields by sandwich waves has been studied extensively. It has been found that for a variety of test fields, the energy of the scattered waves is amplified. In this paper, Scattering of test gravitational waves by sandwich waves are calculated by solving gravitational perturbations in the sandwich wave background. Dependence of their energy amplification with respect to the sandwich wave parameters and the incident wave parameters are examined. The results show that in some cases, the energy of the outgoing test gravitational wave is amplified as well.
\end{abstract}
\section{Introduction}
  Pp waves spacetime is the simplest exact solution of Einstein's gravitational field equation for gravitational waves, which is the spacetime geometry used to characterize the propagation of gravity (mixed with a series of massless particles) along parallel rays[1][2]. Due to the exactness and simplicity of pp waves, it is often used to demonstrate various non-perturbative properties of gravitational waves or as a geometric model of the target space in string theory[3][4][5]. Moreover,  pp waves has universal applicability since any spacetime can be approximated as a pp-wave in the view of an observer moving along the geodesic line close enough to the speed of light, namely Penrose limit[6]. This has led to speculation that some physical phenomena in common spacetime (e.g., many phenomena in the context of the blackholes singularities[7]) can find analogues in pp-wave spacetime.
 \\\\
  One of the most studied subclasses in the pp waves family is the sandwich wave spacetime. It has a planar wave front and the curved part is confined to a limited strip of spacetime, while both sides of the strip are flat spacetime. Therefore, it can be viewed as a kind of "gravitational wave pulse". The optical focusing properties of sandwich waves are the most striking among its many properties. Penrose was the first to discover this property, finding that light emitted from a point in the wavefront region of a sandwich wave focus into a line or a point after passing through the gravitational wave, depending on whether the sandwich wave contains an electromagnetic component[8]. Later, Bond and Pirani generalized this property by considering a family of test particles located in the wave front and placed along some fixed direction. They found that for a sandwich wave with fixed polarization, these test particles collide in finite time after crossing the sandwich wave, regardless of their initial transverse spacing[9].
  \\\\
  After that, M. Halilsoy began to examine the wave version of the focus nature of sandwich waves. Starting from a family of test scalar waves, he found that focusing occurs when this family of test scalar waves passes through a sandwich wave, which is consistent with the case of test particles. He also found that when the parameters of the sandwich wave and the four-momentum of the incident wave satisfy certain conditions, the energy of the outgoing wave is amplified compared to that of the incident wave[10]. He speculated that this should be similar to the superradiation of the test field in Kerr background. Since there is a dependence of the superradiation of the Kerr black hole on the incident wave momentum and spin. Therefore this analogy needs to be examined in a broader sense. Later, Halilsoy examined the massless Dirac field, the electromagnetic field and the N-Abelian field and found a similar situation[11]. However, the shape of the energy amplification factor in the above cases is different from that of the scalar wave.
Immediately after, A. Al-Badawi and M. Kheare abu Shayeb examined the case of charged with mass Dirac field and examined the dependence of the amplification factor on the charged amount and mass[12]. Later, Tekin Dereli, Ozay Gurtug, Mustafa Halilsoy, Yorgo Senikoglu examined the neutrino incident waves case[13]. Afterwards, Yorgo Senikoglu examined the scattering of sandwich waves on the complex gravitino test field[14]. 
\\\\
In the many cases above, the dependence of the amplification factor on the sandwich wave parameters and the incident particle spin and polarization has been carefully examined. However, the case of gravitational test fields are not being considered due to the complexity of linearizing the Einstein gravitational field equations in curved spacetime. In this paper, linearized Einstein gravitational field equations in the sandwich wave background are solved using the Debye potential method developed by Wald, (Wald procedure)[15] and later by Aleksandr Kulitskii and Elena Yu. Melkumova[16]. The obtained solution can be viewed as the scattering of perturbative gravitational waves in a sandwich wave background. Using this solution, energy amplification of the outgoing perturbative gravitational waves is calculated and their properties are analyzed.
\\\\
The chapters of this paper are as follows. Chapter 1 is an introduction. Chapter 2 introduces the linearized Einstein field equation on curved spacetime and the Debye potential method. Chapter 3 introduces sandwich waves and their fundamental properties. Chapter 4 solves the linearized Einstein equation in the sandwich wave context. Chapter 5 calculates the energy amplification from the obtained solutions and analyzes its properties. Chapter 6 is a conclusion.
\section{Linearized Einstein equation}
\subsection{Lichnerowicz equation for spin-2 field}
The linearized Einstein equation is the basic equation of the weak field approximation of general relativity, which assumes that our spacetime metric $g_{\mu\nu}$ can be decomposed into a background metric $^{0}g_{\mu\nu}$ and a perturbation $h_{\mu\nu}$ above it where $h_{\mu\nu}$ is relatively small compared to $^{0}g_{\mu\nu}$, i.e.
\begin{equation}
\begin{split}
g_{\mu\nu}=g^0_{\mu\nu}+h_{\mu\nu},\quad |h^{\mu\nu}| \ll 1.
\end{split}
\end{equation}
Taking the above decomposition into the Einstein field equation and simplify it by neglecting the higher order terms of $h$, making variable substitutions
\begin{equation}
\begin{split}
\psi_{\mu\nu}=h_{\mu\nu}-\frac{1}{2}g_{\mu\nu}h\Leftrightarrow h_{\mu\nu}=\psi_{\mu\nu}-\frac{1}{2}g_{\mu\nu}\psi,\quad \psi=g^{\mu\nu}\psi_{\mu\nu},\quad h=g^{\mu\nu}h_{\mu\nu}.
\end{split}
\end{equation}
After apply Lorentz gauge 
\begin{equation}
\begin{split}
\nabla^{\mu}\psi_{\mu\nu}=0
\end{split}
\end{equation}
to $\psi$, one get the linearized Einstein field equation, namely the Lichnerowicz equation for spin-2 field in curved spacetime.
\begin{equation}
\begin{split}
\nabla^{\alpha}\nabla_{\alpha}\psi_{\mu\nu} + 2R^{\sigma \ \tau}_{\ \mu \ \nu }\psi_{\sigma\tau}-2R^{\sigma}_{\ (\nu }\psi_{\mu)\sigma}=0.
\end{split}
\end{equation}
 This equation can be understood as the equation of motion for the propagation of gravitons in curved spacetime.
\subsection{Wald procedure on pp wave spacetime}
Wald developed a nice procedure for solving the equations of motion for high spin fields in curved spacetime. The equation of motion of a general high spin field has the following form 
\begin{equation}
\begin{split}
\mathscr{E}(f)=0,
\end{split}
\end{equation}
where $f$ is the $n\times 1$ componets field and $\mathscr{E}$ is an matrice linear differential operator with $m\times n$ componets. $\mathscr{E}$ acting on $f$ to get $m$ scalar equations of $f$. Solving these equations directly is difficult. Wald turns to find the matrice linear differential operator with $1\times n$ componets $\mathscr{T}$ such that the equation with respect to $\phi=\mathscr{T}(f)$ can be decoupled i.e. to find differential operator $\mathscr{S},\mathscr{O}$ such that 
\begin{equation}
\begin{split}
\mathscr{S}\mathscr{E}(f)=\mathscr{O}\mathscr{T}(f)=\mathscr{O}(\phi).
\end{split}
\end{equation}
Once these operators are found, the adjoint of the operators can be defined with respect to the corresponding appropriately chosen inner product of fields. Finally, it can be verified that if a scalar field $\xi$ satisfied 
\begin{equation}
\begin{split}
\mathscr{O}^{\dagger}\xi =0,
\end{split}
\end{equation}
then 
\begin{equation}
\begin{split}
f=\mathscr{S}^{\dagger}\xi 
\end{split}
\end{equation}
is a solution to the the equation $\mathscr{E}(f)=0$. The scalar function $\xi$ used to generate the solution $f$ is called the Debye potential and the equation $\mathscr{O}^{\dagger} (\xi) =0$ the Debye potential equation.
\\\\
Kulitskii and Melkumova apply this procedure to gravitational perturbations in the pp waves background, i.e. the metric
\begin{equation}
\begin{split}
ds^2=2dUdV-dX^2 -dY^2 -2H(U,X,Y)dU^2
\end{split}
\end{equation}
in Brinkmann coordinate $(U,V,X,Y)$ where $H(U,X,Y)$ is an arbitrary function of $U,X,Y$.
They found that in pp waves background after taking
\begin{equation}
\begin{split}
\mathscr{O}^{\dagger}=\nabla^{\alpha}\nabla_{\alpha},\quad \mathscr{S}^{\dagger}=2l^{(\mu}\overline{m}^{\nu )}\overline{\delta}D-\overline{m}^{\mu}\overline{m}^{\nu}D^2 -l^{\mu}l^{\nu}\overline{\delta}^2,
\end{split}
\end{equation}
$\psi^{\mu\nu}=\mathscr{S}^{\dagger}\xi$ solves the Lichnerowicz equation for spin-2 field in pp waves spacetime exactly when $\mathscr{O}^{\dagger}\xi =0$ where $l,n,m,\overline{m}$ are Newman-Penrose null tetrads chosen to be
\begin{equation}
\begin{split}
l^{\mu}=\partial_V \quad n^{\mu}=\partial_U +H\partial_V \quad m^{\mu}=\partial_X +\frac{i}{\sqrt{2}}\partial_Y \quad \overline{m}^{\mu}=\partial_X -\frac{i}{\sqrt{2}}\partial_Y 
\end{split}
\end{equation}
and $D, \Delta, \delta ,\overline{\delta}$ are the corresponding Newman-Penrose covariant derivatives 
\begin{equation}
\begin{split}
D =l^{\mu}\nabla_{\mu},\quad \Delta =n^{\mu}\nabla_{\mu},\quad \delta =m^{\mu}\nabla,\quad \overline{\delta} =\overline{m}^{\mu}\nabla_{\mu}
\end{split}
\end{equation}
in pp wave respectively.
\section{Sandwich wave spacetime}
\subsection{Metric}
The sandwich wave spacetime is a type of pp-wave. In Brinkmann coordinates $(U,V,X,Y)$, Its metric can be expressed as
\begin{equation}
\begin{split}
ds^2 = 2dUdV-dX^2 -dY^2 -2H(U,X,Y)dU^2
\end{split}
\end{equation}
where
\begin{equation}
\begin{split}
H(U,X,Y)=-\frac{1}{2} [\Theta (U) - \Theta (U-U_0) ][a^2 (X^2 +Y^2) -b^2 (X^2 -Y^2)],\quad a,b,U_0>0
\end{split}
\end{equation}
and $\Theta (U)$ is the Heavside function.
It is easy to see that $\partial_V$ is a null Killing vector field, which means that the spacetime describes a gravitational wave propagating along the direction of $\partial_V$ at the speed of light. It can be seen that the curved part of the spacetime is only confined to $0<U<U_0$, while the parts of $U<0$ and $U>U_0$ are flat. This suggests that the spacetime describes a gravitational pulse with a phase width of $U_0$. Further, the NP-Weyl scalar and NP-Ricci scalar in the tetrad we mentioned in chapter 2 can be calculated for the sandwich wave spacetime at $0<U<U_0$, , and find that their non-zero components are
\begin{equation}
\begin{split}
\Psi_{4} = -b^2,\quad \Phi_{22}=a^2.
\end{split}
\end{equation}
This means that the coefficients $a,b$ characterize the electromagnetic component and the gravitational component of that gravitational wave, respectively. When $a=0$, the spacetime describes a pure gravitational radiation; when $a\ne 0$, the spacetime describes a gravitational radiation mixed with electromagnetic radiation.
\subsection{Rosen coordinates }
In practice, It is more convenient to use Rosen coordinates instead of Brinkmann coordinates. The relationship between Rosen and Brinkmann coordinates is
\begin{equation}
\left\{
\begin{aligned}
&U = u\quad (u_0 =U_0)\\
&V= v + \frac{1}{2} (x^2 F(u)F'(u) + y^2 G(u)G'(u) )\\
&X=xF(u)\\
&Y=yG(u)
\end{aligned}
\right.
\end{equation}
where
\begin{equation}
\left\{
\begin{aligned}
F(u)=cos[A(u\Theta (u) - (u-u_0)\Theta (u-u_0))]-A\sin(Au_0)(u-u_0)\Theta (u-u_0) \\
G(u)=cos[B(u\Theta (u) - (u-u_0)\Theta (u-u_0))]-B\sin(Bu_0)(u-u_0)\Theta (u-u_0)
\end{aligned}
\right.
\end{equation}
and
\begin{equation}
\begin{split}
A=\sqrt{a^2 -b^2},\quad B=\sqrt{a^2 +b^2}.
\end{split}
\end{equation}
In Rosen coordinates, the spacetime metric can be expressed as
\begin{equation}
\begin{split}
ds^2 =2dudv -F^2 (u)dx^2 - G^2 (u) dy^2.
\end{split}
\end{equation}
It is worth noting that whenever $a\geq b$ or not $F(u)$ and $G(u)$ are real. Zero point of $F(u)$ and $G(u)$ can be found as: 
\begin{equation}
\begin{split}
u_F = u_0 + \frac{1}{A}\cot Au_0,\quad u_G = u_0 + \frac{1}{B}\cot Bu_0.
\end{split}
\end{equation}
Another point worth noting is that the Rosen coordinates do not cover the entire Brinkmann coordinates. Since $F,G=0$ when $u=u_F,u_G$, when $u_F \ne u_G$, the spacetime point with whose Brinkmann coordinates are $(u_F,V,X,Y),X\ne 0$ or $(u_G,V,X,Y),Y\ne 0$ do not correspond to any Rosen coordinates; while when $u_F = u_G$ the spacetime points with whose Brinkmann coordinates $(u_F=u_G,V,X,Y),X \ or \ Y\ne 0$ do not correspond to any Rosen coordinates.
\\\\
The physical meaning of Rosen coordinates is clear: it denotes a geodesic reference system consisting of null observers with initial 4-velocities of $\partial_V$ from the $U<0$ region. Consider a null geodesic observer passing through the point $(0,v_0,x_0,y_0)$ in Brinkmann coordinates, if his affine parameter at the time of passing through the point is zero, then the Rosen coordinates of the spacetime point he is at when his affine parameter is $u_0$ are specified as $(u_0,v_0,x_0,y_0)$. Under this interpretation, the inability of Rosen coordinates to cover the entire Brinkmann coordinates can be interpreted as null observers focusing at $U=u_F\ or \ u_G$: null geodesic observers passing through $(0,v_0,x_0,y_0)$ will always collide in $u_F \ or \ u_G$ phase if they have the same $v_0$, no matter how distance their initial $x_0,y_0$ difference is.
\section{Solution of gravitational perturbation}
The appropriate Debye potential in Rosen coordinates is obtained first. after then it is transferred to Brinkmann coordinates and Wald procedure is being used to obtain the solution of the gravitational perturbation. 
\\\\
Considering a monochromatic scalar plane wave $\xi$ incident from the $U<0$ region traveling against the sandwich wave, which scatters after passing through the sandwich wave. Solving for $\xi$ in Rosen coordinates. Since $\partial_v ,\partial_x ,\partial_y$ are Killing vector fields, it can be assumed that $\xi$ has the following form
\begin{equation}
\begin{split}
\xi =\frac{1}{\sqrt{F(u)G(u)}}\cos (K(u)+\alpha v + p_1 x + p_2 y).
\end{split}
\end{equation}
The Debye potential equation on $\xi$ that we need to solve now is nothing but the D'Alembert equation on the curved spacetime. 
\\\\
\begin{equation}
\begin{split}
\nabla^{\alpha}\nabla_{\alpha}\xi =0.
\end{split}
\end{equation}
Notice that the D'Alembert operator in Rosen coordinates is 
\begin{equation}
\begin{split}
\nabla^\mu \nabla_\mu &= \frac{1}{\sqrt{g}}\partial_\mu (\sqrt{g}g^{\mu\nu} \partial_\nu ) \\
&=2\partial_u \partial_v -\frac{1}{F^2 (u)}\partial_x^2 -\frac{1}{G^2 (u)}\partial_y^2+(\frac{F'(u)}{F(u)}+\frac{G'(u)}{G(u)})\partial_v.
\end{split}
\end{equation}
Substituting the expression of the D'Alembert operator above, one can obtain the equation for $K(u)$ 
\begin{equation}
\begin{split}
2\alpha K'(u) = (\frac{p_1}{F(u)})^2 +  (\frac{p_2}{G(u)})^2.
\end{split}
\end{equation}
Integrating both sides of the equation lead to
\begin{equation}
2\alpha K(u)= \left\{
\begin{aligned}
&(p_1^2 + p_2^2)u+\alpha_0 &u\leq 0\\
&\frac{p_1^2}{A}\tan (Au) + \frac{p_2^2}{B}\tan (Bu) +\alpha_1 &0<u\leq u_0\\
&\frac{p_1^2 \csc^2 (Au_0)}{A^2 (u_F - u )}+\frac{p_2^2 \csc^2 (Bu_0)}{B^2 (u_G - u )}+\beta &u>u_0
\end{aligned}
\right.
\end{equation}
where $\alpha_0, \alpha_1, \beta$ are arbitrary constant for integration. Since what is calculated is a monochromatic plane wave incident from the $u<0$ region, it can be set that $\alpha_0 =0$. From the continuous condition we have
\begin{equation}
\begin{split}
\alpha_1 = 0,\quad \beta =-\frac{p_1^2}{A}\cot (Au_0) -\frac{p_1^2}{B}\cot (Bu_0).
\end{split}
\end{equation}
Thus
\begin{equation}
K(u)= \left\{
\begin{aligned}
&\frac{p_1^2 + p_2^2}{2 \alpha}u &u\leq 0\\
&\frac{1}{2 \alpha}[\frac{p_1^2}{A}\tan (Au) + \frac{p_2^2}{B}\tan (Bu)] &0<u\leq u_0\\
&\frac{p_1^2 \csc^2 (Au_0)}{A^2 (u_F - u )}+\frac{p_2^2 \csc^2 (Bu_0)}{B^2 (u_G - u )}-\frac{p_1^2}{A}\cot (Au_0) -\frac{p_1^2}{B}\cot (Bu_0)] &u>u_0.
\end{aligned}
\right.
\end{equation}
Transform back to Brinkmann coordinates one get
\begin{equation}
\begin{split}
\xi &=\frac{1}{\sqrt{F(U)G(U)}}\cos (K(U)+\alpha V -\frac{1}{2}X^2 \frac{F'(U)}{F(U)}-\frac{1}{2}Y^2 \frac{G'(U)}{G(U)}+p_1 \frac{X}{F(U)}+p_2 \frac{Y}{G(U)})\\
&=\frac{1}{\sqrt{F(U)G(U)}}\cos S(U,V,X,Y)\\
\end{split}
\end{equation}
where we denote $S(U,V,X,Y)=K(U)+\alpha V -\frac{1}{2}X^2 \frac{F'(U)}{F(U)}-\frac{1}{2}Y^2 \frac{G'(U)}{G(U)}+p_1 \frac{X}{F(U)}+p_2 \frac{Y}{G(U)}$.
\\\\
Substituting $\xi$ into $\psi =\mathscr{S}^{\dagger}\xi$ we have
\begin{equation}
\begin{split}
\psi^{\mu\nu}=H^{\mu\nu}\frac{1}{\sqrt{F(U)G(U)}}\cos S(U,V,X,Y)
\end{split}
\end{equation}
where $H^{\mu\nu}$ is a $4\times 4$ traceless symmetric complex matrix whose non-zero components are
\begin{equation}
\begin{split}
H^{V V}&=\frac{\alpha}{2}(\frac{G'(U)}{G(U)}-\frac{F'(U)}{F(U)})\tan S(U,V,X,Y)+\frac{1}{2}(\frac{p_1-\alpha X F'(U)}{F(U)}-i\frac{p_2 -\alpha Y G'(U)}{G(U)})^2,\\
H^{V X}&=H^{X V}=-\frac{\alpha}{2}(\frac{p_1-\alpha X F'(U)}{F(U)}-i\frac{p_2 -\alpha Y G'(U)}{G(U)}),\\
H^{V Y}&=H^{Y V}=\frac{i\alpha}{2}(\frac{p_1-\alpha X F'(U)}{F(U)}-i\frac{p_2 -\alpha Y G'(U)}{G(U)}),\\
H^{X X}&=\frac{1}{2}\alpha^2,\\
H^{X Y}&=H^{Y X}=-\frac{i}{2}\alpha^2,\\
H^{Y Y}&=-\frac{1}{2}\alpha^2.\\
\end{split}
\end{equation}
It can be seen that $H^{\mu\nu}$ is a complex tensor field, while its real and imaginary parts represent the two polarizations of gravitational waves, respectively. 
\section{Energy analysis}
This chapter calculate the energy change of the gravitational perturbation after being scattered by the sandwich wave. Since what is examined is the change in the energy of the gravitational perturbation before and after scattering, and in this case, the gravitational perturbation is on the flat spacetime (i.e. $U<0$ and $U>U_0$ regions), the stress-energy formula of the weak gravitational field in the flat spacetime can be used to perform calculation[18], namely
\begin{equation}
\begin{split}
T^{\mu\nu}=\frac{1}{4} [2\psi^{\alpha\beta, \mu} \psi_{\alpha\beta}^{\ \ \ ,\nu}-\psi^{,\mu}\psi^{,\nu}-\eta^{\mu\nu} (\psi^{\alpha\beta, \sigma}\psi_{\alpha\beta, \sigma}-\frac{1}{2}\psi^{,\sigma}\psi_{,\sigma})].
 \end{split}
\end{equation}
On the other hand, the energy density $T^{00}$ in Cartesian coordinates can be expressed in terms of the components of the stress capacity tensor in light-cone coordinates, i.e.
\begin{equation}
\begin{split}
T^{00}=\frac{1}{2}(T^{UU}+2T^{UV}+T^{VV}).
 \end{split}
\end{equation}
Firstly, the real part of $\psi^{\mu\nu}$ can be taken and substitute into the expression of the stress-energy tensor. According to the above formula, the expression of $T^{00}$ can be obtained.  Turther, restricting $T^{00}$ to $U=0$ and $U=U_0$ respectively yields
\begin{equation}
\begin{split}
T^{00}|_{U=0}&=\frac{\alpha^2}{32}(2\alpha^2 +p_1^2+p_2^2)^2 \sin^2 (\alpha V +p_1 X +p_2 Y])\quad  (U<0),\\
T^{00}|_{U=U_0}&=\frac{1}{32}\alpha^4 \{(\frac{4 \alpha^2}{c_A c_B} + \frac{4(p_1 +\alpha A X s_A)^2}{c_A^3 c_B} + \frac{4(p_2 +\alpha B Y s_B)^2}{c_A c_B^3}) \sin^2 S(U_0,V,X,Y)\\
&+[-\frac{A c_B s_A+B c_A s_B}{c_A^\frac{3}{2} c_B^\frac{3}{2}}\cos S(U_0,V,X,Y) +(\frac{p_1^2 c_A^{-2}+ p_2^2 c_B^{-2}}{\alpha c_A^\frac{1}{2}c_B^\frac{1}{2}}\\
& +\frac{(2p_1 +\alpha A X s_A) A X  s_A}{c_A^\frac{5}{2}c_B^\frac{1}{2}}+\frac{(2p_2 +\alpha B Y s_B) B Y s_B}{c_A^\frac{1}{2}c_B^\frac{5}{2}})\sin S(U_0,V,X,Y)]^2\}\\ 
\end{split}
\end{equation}
where $c_A =\cos (AU_0), c_B =\cos (BU_0), s_A =\sin (AU_0), s_B =\sin (BU_0)$.
\\\\
Differentiating the above two equations, one get the energy amplification
\begin{equation}
\begin{split}
\Delta T^{00}&= T^{00}|_{U=U_0}-T^{00}|_{U=0}\\
&=\frac{1}{32}\alpha^4 \{(\frac{4 \alpha^2}{c_A c_B} + \frac{4(p_1 +\alpha A X s_A)^2}{c_A^3 c_B} + \frac{4(p_2 +\alpha B Y s_B)^2}{c_A c_B^3}) \sin^2 S(U_0,V,X,Y)\\
&+[-\frac{A c_B s_A+B c_A s_B}{c_A^\frac{3}{2} c_B^\frac{3}{2}}\cos S(U_0,V,X,Y) +(\frac{p_1^2 c_A^{-2}+ p_2^2 c_B^{-2}}{\alpha c_A^\frac{1}{2}c_B^\frac{1}{2}}\\
& +\frac{(2p_1 +\alpha A X s_A)A X s_A}{c_A^\frac{5}{2}c_B^\frac{1}{2}}+\frac{(2p_2 +\alpha B Y s_B) B Y s_B}{c_A^\frac{1}{2}c_B^\frac{5}{2}})\sin S(U_0,V,X,Y)]^2\}\\ 
&-\frac{\alpha^2}{32}(2\alpha^2 +p_1^2+p_2^2)^2 \sin^2 (\alpha V +p_1 X +p_2 Y).
\end{split}
\end{equation}
Averaging by $\Delta \overline{T}^{00}=\lim_{\tau \rightarrow \infty}\frac{1}{\tau }\int^{\tau /2}_{-\tau /2} \Delta T^{00} {\rm d}V$ one obtains
\begin{equation}
\begin{split}
\Delta \overline{T}^{00}&=\frac{1}{64}\alpha^4 [\frac{4 \alpha^2}{c_A c_B} + \frac{4(p_1 +  \alpha A X s_A)^2}{c_A^3 c_B} + \frac{4(p_2 +\alpha B Y s_B)^2}{c_A c_B^3}+(\frac{A c_B s_A+B c_A s_B}{c_A^\frac{3}{2} c_B^\frac{3}{2}})^2 \\
&+(\frac{p_1^2 c_A^{-2}+ p_2^2 c_B^{-2}}{\alpha c_A^\frac{1}{2}c_B^\frac{1}{2}} +\frac{(2p_1 +\alpha A X s_A) A X s_A}{c_A^\frac{5}{2}c_B^\frac{1}{2}}+\frac{(2p_2 +\alpha B Y s_B)  B Y s_B}{c_A^\frac{1}{2}c_B^\frac{5}{2}})^2 -\frac{(2\alpha^2 +p_1^2+p_2^2)^2}{\alpha^2}].
\end{split}
\end{equation}
Calculation of the energy amplification of the imaginary part is exactly the same way, surprisingly, the result is exactly the same as for the real part.
 \\\\
From the above expression, the following properties can be seen:
\\\\
1. The expression is symmetric with respect to $(X,p_1,F)\leftrightarrow (Y,p_2,G)$;\\
2. When $U_0\rightarrow 0$ or $a,b\rightarrow 0$, $\Delta T^{00} \rightarrow 0$; \\
3. $\Delta \overline{T}^{00}$ does not depend on the polarization of the incident gravitational waves.\\
4. When $\cos (AU_0)\rightarrow 0$ or $\cos (BU_0)\rightarrow 0$, $u_F \rightarrow U_0$ or $u_G \rightarrow U_0$, $\Delta \overline{T}^{00} \rightarrow +\infty$, that is, in both cases, hypersurface $U=U_0$ and the focusing hypersurface coincide, where the energy amplification is infinite.\\
5. Fixing $a,b,U_0$, for sufficiently large $X,Y$, there is always $\Delta \overline{T}^{00}>0$, i.e., an amplification in the energy of the outgoing wave occurs.
\\\\
All five of these properties are consistent with the other test fields previously calculated.
\section{Conclusion}
The solution of the gravitational perturbation in the sandwich wave background is calculated by the Wald procedure, which can be interpreted as the scattering of gravitons in the sandwich wave. The results show that the amplification of the outgoing wave energy does not depend on the polarization of the incident wave; the amplification of the outgoing wave energy tends to infinity when the parameters of the sandwich wave satisfy certain conditions so that the outgoing surface tends to coincide with the focusing hypersurface; when the parameters of the sandwich wave are fixed, the energy amplification always occurs where $X,Y$ is sufficiently large, regardless of the 4-momentum of the incident wave. It can be found that these properties are the same as those of the other test fields that have been studied previously.

\bibliographystyle{alpha}
\bibliography{sample}

\end{document}